\input{aipcheck}

\documentclass[
    ,final            % use final for the camera ready runs
  ]
  {aipproc}

\layoutstyle{8x11double}

\begin{document}

\title{The very first Pop III stars and their relation to bright $z \approx 6$ quasars}


\author{Michele Trenti}{
 address={Space Telescope Science Institute, 3700 San Martin Drive Baltimore MD 21218 USA}
}

\author{Massimo Stiavelli}{
  address={Space Telescope Science Institute, 3700
San Martin Drive Baltimore MD 21218 USA},altaddress={Department of Physics and
Astronomy, Johns Hopkins University, Baltimore, MD 21218 USA}
}

\begin{abstract}

We discuss the link between dark matter halos hosting the first PopIII
stars formed at redshift $z > 40$ and the rare, massive, halos that
are generally considered to host bright $z \approx 6$ quasars. We show
that within the typical volume occupied by one bright high-z QSO the
remnants of the first several thousands PopIII stars formed do not end
up in the most massive halos at $z \approx 6$, but rather live in a
large variety of environments. The black hole seeds planted by these
very first PopIII stars can easily grow to $M > 10^{9.5} M_{\odot}$ by
$z=6$ assuming Eddington accretion with radiative efficiency $\epsilon
\approx 0.1$. Therefore quenching of the accretion is crucial to avoid
an overabundance of supermassive black holes. We implement a simple
feedback model for the growth of the seeds planted by PopIII stars and
obtain a $z \approx 6$ BH mass function consistent with the observed QSO
luminosity function.

\end{abstract}

\maketitle

%%%%%%%%%%%%%%%%%%%%%%%%%%%%%%%%%%%%%%%%%%%%
%% MAINMATTER
%%%%%%%%%%%%%%%%%%%%%%%%%%%%%%%%%%%%%%%%%%%%

\section{Introduction}

Population III stars formed in the early universe at redshift $z>20$
with a top-heavy initial mass function (e.g. see \cite{bro04}) are
expected to leave at the end of their lives intermediate mass black
remnants of the order of $100 M_{\odot}$. These seeds, formed within
dark matter halos of mass $\approx 10^6 M_{\odot}$, may be the
starting points for accretion that will lead to supermassive black
holes ($M_{BH}>10^9 M_{\odot}$), which are considered to power the
luminosity of QSOs, observed in the Sloan Digital Sky Survey (SDSS) at
$z>6$ when the universe was less than one billion years old (e.g. see
\cite{fan04}). These bright QSOs are extremely rare objects (one
object per about 200 deg$^2$ in SDSS, see \cite{fan04}), so we expect
on average one per 1Gpc$^3$ comoving. 

Within this volume the QSO may either be the descendant of the first
intermediate mass black hole seed left from the \emph{first} PopIII
star, which would therefore give the most time for mass accretion, or
sit at the center of the most massive structure at $z \approx 6$. Of
course these two alternatives are in principle mutually non-exclusive,
as the remnants of the first PopIII stars could end up in the most
massive dark matter halos at $z\approx 6$. This possibility seems to
be implied by a number of recent papers, where the progenitor halos of
bright quasars are traced back in time and identified as the first
dark matter halos formed in the universe (e.g. see \cite{MR},
\cite{ree05}, \cite{li06}). 
However these works either do not have the mass resolution to identify
the dark matters halos hosting the first generation of PopIII stars or
rely on multiple mesh refinements of a small region centered around
the largest halos identified at z=0 in order to resolve scales down to
$10^6 M_{\odot}$. 

To properly address the link between bright quasars and PopIII stars
it is necessary to resolve a dynamic range in mass of more than
$10^{13}$: a simulation box of 1 Gpc$^3$ contains a mass larger than
$10^{19} M_{\odot}$ and within this box dark matter halos below $10^6
M_{\odot}$ need to be identified. Therefore we have adopted an
original approach (see \cite{tre07}), broadly based on the tree method
by \cite{col94}. The idea is based on coupling a numerical simulations
of structure formation to identify dark matter halos at $z
\approx 6$ with a Monte Carlo method to sample subgrid fluctuations of the initial
Gaussian random field of density fluctuations at the mass scale
typical of halos hosting PopIII. This allows us to attach to every
particle in the simulation, which has typically a mass in excess of
$10^{10} M_{\odot}$, the formation time of its first PopIII star
progenitor. The details of the method are presented in detail in
\cite{tre07} along with an extensive set of tests to validate our
innovative approach. Here we focus instead on summarizing the main
results from our investigation.

In particular we show that the first PopIII progenitor of a $z=6$
bright QSO, while born at $z>40$ well within the early era of PopIII
formation, it is not the first PopIII in the typical Gpc$^3$ volume
occupied by the QSO, but it is rather preceded by about $10^4$ other
PopIII stars. A qualitative understanding can be reached from
simple considerations based on the properties of Gaussian random
fields deriving from the spectrum of primordial density perturbations:
small mass dark matter halos are sensitive to higher frequency in the
density fluctuations spectrum than their higher mass
counterparts. Therefore the first $10^6 M_{\odot}$ dark matter halos
formed at $z>45$ in a simulation box will not in general evolve to
become the first $10^{12} M_{\odot}$ dark matter halos formed at
$z\approx 6$.

In terms of intermediate mass black hole growth from PopIII this
result implies that there are a number of seeds formed in the early
universe before the one that will become the bright z=6 QSO. All these
seeds have enough time, if accreting at Eddington limit with accretion
efficiency $\epsilon \approx 0.1$ to become supermassive ($M_{BH} >
10^9 M_{\odot}$) by z=6. We follow their evolution and we show with a
simple accretion model that the gas supply available for growth is
limited for most of these seeds, so that the QSO luminosity function
derived in our framework is consistent with the slope of the observed
QSO luminosity function.

%%%%%%%%%%%%%%%%%%%%%%%%%%%%%%%%%%%%%%%%%%%%%
\begin{figure} 
  \includegraphics[height=.3\textheight]{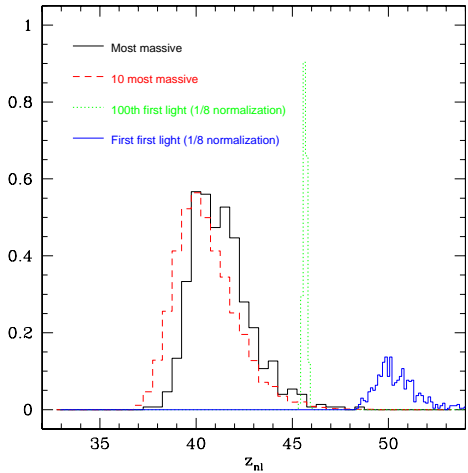}
  \caption{From \cite{tre07}: Distribution of the collapse time $z_{nl}$ for the oldest
  PopIII progenitor (with $M_{fs} = 10^6 M_{\odot}/h$) of the most
  massive halo (black line) and averaged over the 10 most massive
  halos (red line) at $z=6$ in the $(720 Mpc/h)^3$ box simulation. The
  blue line represents the collapse redshift of the \emph{first}
  PopIII star perturbation, while the dotted green line refers to the
  collapse redshift of the 100th PopIII in the
  box.}\label{fig:red720}
\end{figure}
%%%%%%%%%%%%%%%%%%%%%%%%%%%%%%%%%%%%%%%%%%%%%%%%

\section{Numerical simulations}

We identify the largest dark matter halos at $z=6$ in three
cosmological simulations with $512^3$ particles and different box
sizes: a large (edge $720$ Mpc/h), a medium (edge $512$ Mpc/h) and a
small (edge $60$ Mpc/h) box. The simulations have been carried out
with the public version of the tree-PM code Gadget2 \cite{spr05} and a
cosmology based on third year WMAP data \cite{WMAP3}:
$\Omega_{\Lambda}=0.74$, $\Omega_{m}=0.26$, $H_0=70~km/s/Mpc$, where
$\Omega_m$ is the total matter density in units of the critical
density ($\rho_{c}= 3H_0^2/(8
\pi G)$) with $H_0$ being the Hubble constant (parameterized as $H_0 =
100 h~ km/s/Mpc$) and $G$ the Newton's gravitational
constant.. $\Omega_{\Lambda}$ is the dark energy density. In
generating the initial density field we use a scale invariant
long-wave spectral index ($n=1$) of the power spectrum of density
fluctuations and $\sigma_8=0.9$ or $\sigma_8=0.75$ (the root mean
squared mass fluctuation in a sphere of radius $8Mpc/h$ extrapolated
at $z=0$ using linear theory). As described in \cite{tre07}, the
initial density field is then used as input in our Monte Carlo code to
obtain the formation redshift of the first PopIII progenitor of each
particle in the simulation box.

\section{Results}\label{sec:res}

%%%%%%%%%%%%%%%%%%%%%%%%%%%%%%%%%%%%%%%%%%%%%
\begin{figure} 
  \includegraphics[height=.3\textheight]{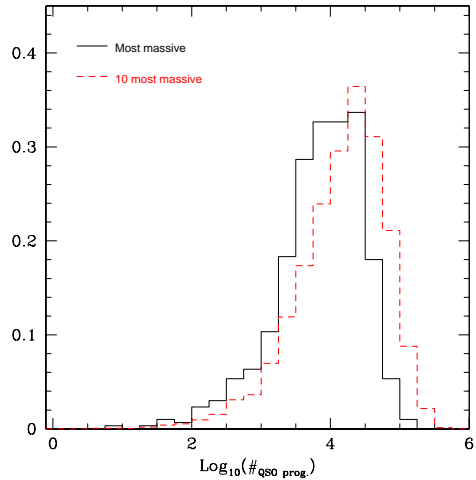} \caption{From
  \cite{tre07}: Distribution for the ranking of the collapse epoch for
  the oldest PopIII halo progenitor of the most massive halo (black
  line) and averaged over the 10 most massive halos (red line) at
  $z=6$ in the $(720 Mpc/h)^3$ box simulation.}\label{fig:card720}
\end{figure}
%%%%%%%%%%%%%%%%%%%%%%%%%%%%%%%%%%%%%%%%%%%%%%%%

\begin{figure} 
  \includegraphics[height=.3\textheight]{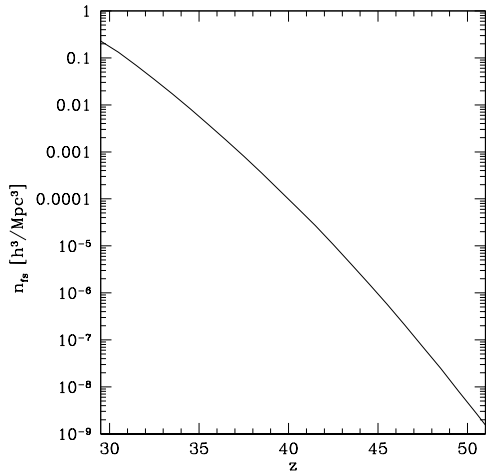}
\caption{From \cite{tre07}: Comoving number density of virialized
  PopIII halos ($M_{fs}=10^6 M_{\odot}/h$) per unit redshift versus
  redshift for $z>29$ as measured by means of our MC
  code.}\label{fig:sfr}
\end{figure}

Under the assumption that the first PopIII stars in the universe have
formed in $10^6 M_{\odot}$ halos cooled by molecular hydrogen, the
progenitor of the most massive $z=6$ halo (the one assumed to host the
bright QSO) in our large box simulation is born at $z
\approx 41$, while the first PopIII in the volume is already present
at $z>49$ (see Fig.~\ref{fig:red720}). By the time the QSO progenitor
is born, there are on average 8000 other PopIII stars formed in the
simulation (see Fig.~\ref{fig:card720}), with a typical PopIII star
formation rate, as obtained from our method, shown in
Fig.~\ref{fig:sfr}. If PopIII stars reside in halos of different mass
and/or if the value of $\sigma_8$ is different, there will be a shift
on the formation time of these first stars (that is on the scale of
the x-axis in Figs.~\ref{fig:red720}-\ref{fig:sfr}), but the relative
ranking between the \emph{first} PopIII star in the box and the PopIII
that is the QSO progenitor remains essentially unchanged (e.g. see
Fig.~3 in \cite{tre07}).

%%%%%%%%%%%%%%%%%%%%%%%%%%%%%%%%%%%%%%%%%%%%%
%%%%%%%%%%%%%%%%%%%%%%%%%%%%%%%%%%%%%%%%%%%%%%%%

\begin{figure} 
  \includegraphics[height=.3\textheight]{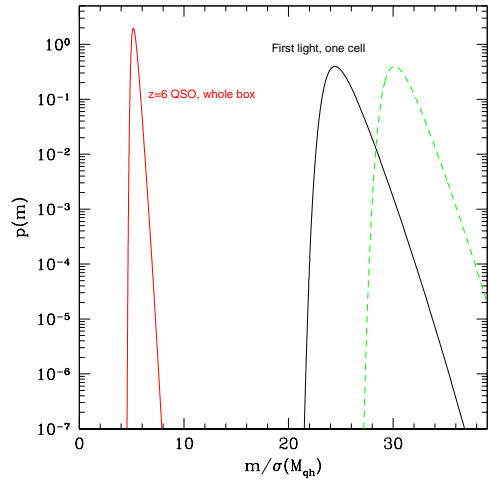} \caption{From
   \cite{tre07}: Probability distribution functions $p(m)$ for the
   maximum $m$ of $k$ random Gaussian fluctuations representative of
   mass scale for halos hosting $z=6$ QSO candidates (red curve) and
   PopIII stars (black curve). To compute $p(m)$ for QSO hosting halos
   we have assumed a box of $720Mpc/h$ and a mass scale of $M_{qh} =
   4.3~10^{12} M_{\odot}/h$ which implies $k=180^3$. This curve
   represents the probability distribution for a sigma peak that at $z
   \approx 6$ leads to one of the most massive halos in the simulation
   volume. The curve associated to first light perturbations (solid
   black, with $M_{fs}=10^{6} M_{\odot}/h$) is derived using
   $\sigma_{fs}^2={\sigma_{M_{fs}}^2-\sigma_{M_{qh}}^2}$ and
   $k=160^3$: it represents the probability distribution for the
   maximum of the sub-grid scale fluctuations at the $M_{fs}$ scale
   within one cell of the $180^3$ volume. The dashed green line
   represents the probability distribution of the density fluctuation
   associated to the first PopIII progenitor of a $5.7 \sigma(M_{qh})$
   peak. $m$ is given in units of $\sigma_{M_{qh}}$.}\label{fig:pdf}
\end{figure}

The relation between PopIII and QSOs can be easily understood in terms
of the statistics of Gaussian random fields (see
Fig.~\ref{fig:pdf}). In fact, the most massive $z=6$ halo in the
simulation box (with mass $M_{qh=}\approx 4.3 \times 10^{12}
M_{\odot}$) originates from a $\approx 6
\sigma(M_{qh})$ peak in the density perturbation field. If we know
consider a random volume in the simulation box with mass $M_{qh}$, we
find that the distribution for the maximum peak at the mass scale of a
PopIII halo ($10^6 M_{\odot}$) inside this volume is in the range
$[23:27]
\sigma(M_{qh})$ at the 90\% confidence level. This is essentially because (i) we are
at a smaller mass scale, so there is additional power in the density
fluctuations field and (ii) we are considering here the first PopIII
halo formed in the $M_{qh}$ volume, that is the maximum among $160^3$
fluctuations. The probability distribution for the peak associated to
the first PopIII progenitor of the QSO (green dashed line in
Fig.~\ref{fig:pdf}) is then given by combining the probability
distribution of the QSO overdensity with that of the first PopIII halo
of a random cell. From Fig.~\ref{fig:pdf} it is immediately apparent
that the advantage of sitting at the top of the QSO overdensity is not
sufficient for the first PopIII in this volume to beat all other
PopIII in the box. Still one in about $10^3$ PopIII progenitors of
random $M_{qh}$ cells in the (720 Mpc/h)$^3$ box is formed before the
QSO progenitor. As there are $\approx 6 \times 10^6$ cells of mass
$M_{qh}$ in this volume, we expect from this simple consideration that
the PopIII progenitor of the QSO will be formed when already $\approx
6000$ other PopIII stars are present in the box, in excellent
agreement with the detailed results from the numerical simulation (see
Fig.~\ref{fig:card720}).

\begin{figure} 
  \includegraphics[height=.3\textheight]{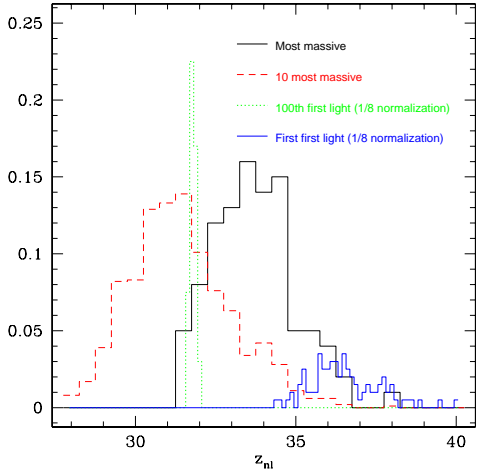}
\caption{From \cite{tre07}: Like in
  Fig.~\ref{fig:red720}, but for a smaller box simulation, with volume
  of $(60 Mpc/h)^3$ and $\sigma_8 = 0.75$. Given the limited volume
  available sitting on the top of the overdensity that will collapse
  to the largest $z=6$ halo gives a significant advantage to PopIII
  halos formed at $z>30$, which are typically within the first 100
  formed in the box.}\label{fig:red60}
\end{figure}

This result is essentially given by the fact that bright QSOs at $z=6$
are very rare objects. If fact, if we consider a smaller cosmic volume
(e.g. a box of edge $60$ Mpc/h) and repeat the experiment of studying
the formation time of the first PopIII progenitor of the most massive
$z=6$ halo, we find different results (see Fig.~\ref{fig:red60}). In
this case the cosmic volume considered is significantly smaller, so
sitting at the top of the most massive $z=6$ halo gives a significant
relative advantage to PopIII formation at $z>30$: the PopIII
progenitor of the most massive halo is typically within the first 10
to 100 PopIII in the box.

%%%%%%%%%%%%%%%%%%%%%%%%%%%%%%%%%%%%%%%%%%%%%
%%%%%%%%%%%%%%%%%%%%%%%%%%%%%%%%%%%%%%%%%%%%%%%%

%%%%%%%%%%%%%%%%%%%%%%
%%%%%%%%%%%%%%%%%%%%%%

\section{IMBH seeds growth}

From our investigation it is clear that, before the first PopIII
progenitor of the most massive halo at $z=6$ is born, several
thousands of intermediate mass ($m_{BH}\approx 10^2{M_{\odot}}$) black
hole seeds are planted by PopIII stars formed in a cosmic volume that
will on average host a bright $z=6$ quasar. Here we investigate with a
simple merger tree code what is the fate of the black holes seeds
formed up to the formation time of the quasar seed and what are the
implications for the observed quasar luminosity function.

We assume Eddington accretion for the BH seeds, so that the evolution
of the BH mass is given by: 
%%%%
\begin{equation}
m_{BH} = m_{0} \exp{\left [(t-t_0)/t_{sal} \right ]},
\end{equation}
%%%%
where $m_0$ is the mass at formation time $t_0$ and $t_{sal}$ is the
Salpeter time \citep{sal64}:
%%%%
\begin{equation} \label{eq:accr}
t_{sal} = \frac{\epsilon ~m_{BH} ~c^2}{(1-\epsilon)L_{Edd}} = 4.507 \cdot
10^{8} yr \frac{\epsilon}{(1-\epsilon)}, 
\end{equation}
%%%%
where $\epsilon$ is the radiative efficiency.

If $\epsilon \approx 0.1$ there has been enough time to build up a
$z\approx 6$ supermassive black hole with mass $m_{BH} > 10^9$
starting from a PopIII remnant formed at $z>40$. This highlights that
only a minor fraction of the PopIII BH seeds formed before $z=40$ can
accrue mass with high efficiency, otherwise the number density of
supermassive black holes at low redshift would greatly exceed the
observational constraints. The first BH seeds in the box are distant
from each other, so they evolve in relative isolation, without
possibly merging among themselves. Therefore other mechanisms must be
responsible for quenching accretion of the first BH
seeds. Interestingly if we were to assume that accretion periods are
Poisson distributed in time for each seed, we would not be able to
explain the observed power law distribution of BH masses at $z <
6$ around the high mass end. A Poisson distribution would in fact give
too little scatter around the median value and a sharp (faster than
exponential) decay of the displacements from the mean accreted
mass. An exponential distribution of the accretion efficiency is
instead required to match the observed BH mass function. In addition,
it is necessary to assume that the duty cycle of the BH accretion is
roughly proportional to the mass of the halo it resides in. To explore
this possibility we follow the merging history of PopIII halos formed
at $z=40$ by means of a merger-tree code. We implement a BH growth
based on Eq.~\ref{eq:accr}, but at each step of the tree we limit the
BH mass to $m_{BH} \leq
\eta~ m_{bar}$, where $m_{bar}$ is the total baryon mass of the halo
that hosts the BH. If $\eta \approx 6 \cdot 10^{-3}$ (like in
\cite{yoo04}), then we obtain an expected
mass for the BH powering bright $z=6$ quasar of $\approx 5
\cdot 10^{9} M_{\odot}$, which is in agreement with the observational
constraints from SDSS quasars \citep{fan04}. By fitting a power law
function to the BH mass function in the range $[0.055:0.2]\cdot
10^{10} M_{\odot}$ we obtain a slope $\alpha \approx -2.6$, while the
slope is $\alpha \approx -3.7$ in the mass range $[0.2:1.0] \cdot
10^{10} M_{\odot}$, a value that is consistent within the $1 \sigma$
error bar with the slope of the bright end of the quasar luminosity
function measured by \cite{fan04}.

\section{Conclusion}\label{sec:con}

We study the link between the first PopIII halos collapsed in a
simulation box and the most massive structures at $z
\approx 6$, with the aim of establishing the relationship between the
first intermediate mass black holes created in the universe and the
super-massive black holes that power the emission of bright $z=6$
quasars. We show that almost no correlation is present between the
sites of formation of the first few hundred $10^6 M_{\odot}/h$ halos
and the most massive halos at $z \leq 6$ when the simulation box
has an edge of several hundred $Mpc$. Here the PopIII progenitors
(halos of mass $M_{fs} \approx 10^6 M_{\odot}$) of massive halos at $z
\leq 6$ formed from density peaks that are $\approx 1.5
\sigma(M_{fs})$ more common than that of the \emph{first} PopIII star
in the (720 Mpc/h)$^3$ simulation box. These halos virialize around
$z_{nl} \approx 40$, to be compared with $z_{nl} \geq 48$ of the
\emph{first} PopIII halo.

This has important consequences. We show that, if bright quasars and
supermassive black holes live in the most massive halos at $z\approx
6$, then their progenitors at the $10^6 M_{\odot}$ mass scale are well
within the PopIII era, regardless of the PopIII termination
mechanism. On the other hand, if the $m_{BH}/\sigma$ relationship is
already in place at $z=6$, then bright quasars are not linked to the
remnants of the very first intermediate mass black holes (IMBHs) born
in the universe, as their IMBH progenitors form when already several
thousands of PopIII stars have been created within the typical volume
that hosts a bright $z=6$ quasar. The IMBH seeds planted by this very
first PopIII stars have sufficient time to grow up to $m_{BH} \in
[0.2:1] \cdot 10^{10} M_{\odot}$ by $z=6$ if we assume Eddington
accretion with radiative efficiency $\epsilon \leq 0.1$. Instead,
quenching of the BH accretion is required for the seeds of those
PopIII stars that will not end up in massive halos at $z=6$, otherwise
the number density of supermassive black holes would greatly exceed
the observational constraints. One way to obtain growth consistent
with the observations is to limit the accreted mass at a fraction
$\eta \approx 6 \cdot 10^{-3}$ of the total baryon halo mass. This
gives a slope of the BH mass function $\alpha = -3.7$ in the BH mass
range $m_{BH} \in [0.2:1] \cdot 10^{10} M_{\odot}$, which is within
the $1 \sigma$ uncertainty of the slope of the bright end of the $z=6$
quasar luminosity function ($\alpha \approx -3.5$) measured by
\cite{fan04}.

Another important point highlighted by this study is that rich
clusters do not preferentially host the remnants of the first PopIII
stars. In fact the remnants of the first 100 Pop-III stars in our
medium sized simulation box (volume of $(512 Mpc/h)^3$) end up at
$z=0$ on halos that have a median mass of $3 \cdot 10^{13}
M_{\odot}/h$. This suggests caution in interpreting the results from
studies that select a specific volume of the simulation box, like a
rich cluster, and then progressively refine smaller and smaller
regions with the aim of hunting for the first lights formed in the
whole simulation (see e.g., \cite{ree05,li06}). Only by considering
refinements over the complete volume of the box the rarity and the
formation ranking of these progenitors can be correctly evaluated.

%%%%%%%%%%%%%%%%%%%%%%%%%%%%%%%%%%%%%%%%%%%%
%% Sample figure:
%%
%% The option [height=...] scales the picture to the given height,
%% without it it would be printed at its nominal size
%%%%%%%%%%%%%%%%%%%%%%%%%%%%%%%%%%%%%%%%%%%%

%%%%%%%%%%%%%%%%%%%%%%%%%%%%%%%%%%%%%%%%%%%%%%%%
%% BACKMATTER
%%%%%%%%%%%%%%%%%%%%%%%%%%%%%%%%%%%%%%%%%%%%%%%%

\begin{theacknowledgments}
This work was supported in part by NASA JWST IDS grant NAG5-12458 and
by STScI-DDRF award D0001.82365.
\end{theacknowledgments}

%%%%%%%%%%%%%%%%%%%%%%%%%%%%%%%%%%%%%%%%%%%%%%%%
%% The bibliography can be prepared using the BibTeX program or
%% manually.
%%
%% The code below assumes that BibTeX is used.  If the bibliography is
%% produced without BibTeX comment out the following lines and see the
%% aipguide.pdf for further information.
%%
%% For your convenience a manually coded example is appended
%% after the \end{document}
%%%%%%%%%%%%%%%%%%%%%%%%%%%%%%%%%%%%%%%%%%%%%%%%

%%%%%%%%%%%%%%%%%%%%%%%%%%%%%%%%%%%%%%%%%%%%%%%%
%% You may have to change the BibTeX style below, depending on your
%% setup or preferences.
%%
%%
%% For The AIP proceedings layouts use either
%%%%%%%%%%%%%%%%%%%%%%%%%%%%%%%%%%%%%%%%%%%%

\bibliographystyle{aipproc}   % if natbib is available

\begin{thebibliography}{99}

\bibitem{bro04} Bromm, V. and Larson, R.~B. 2004, ARA\&A, 42, 79

\bibitem{fan04} {Fan}, X. et al. 2004, AJ, 128, 515

\bibitem{MR} Springel, V. et al. 2005, Nature, 435, 629

\bibitem{ree05} Reed, D.~S. et al. 2005, MNRAS, 363, 393

\bibitem{li06} Li, Y. et al. 2006, ApJ, submitted, astro-ph/0608190

\bibitem{tre07} Trenti, M. and Stiavelli, M. 2007, ApJ, 667, 38

\bibitem{col94} {Cole}, S. et al. 1994, MNRAS, 271, 781

\bibitem{spr05} Springel, V. 2005, MNRAS, 364, 1105

\bibitem{WMAP3} Spergel, D.~N. et al. 2007, ApJS, 170, 377

\bibitem{sal64} Salpeter, E.~E. 1964, ApJ, 140, 796

\bibitem{yoo04} Yoo, J. and Miralda-Escud{\'e}, J. 2004, ApJL, 614, 25

\end{thebibliography}

\end{document}